\documentclass[aps,reprint,prb,showpacs,amsmath,amssymb]{revtex4-1}
\usepackage{graphicx}
\usepackage{hyperref}
\usepackage{longtable}
\usepackage{mathtools}
\usepackage[utf8]{inputenc}
\usepackage[T1]{fontenc}
\begin{document}
\title{Magnetoresistance in LuBi and YBi semimetals\\ due to nearly perfect carrier compensation}
\author{Orest Pavlosiuk$^1$} 
\author{Przemysław Swatek$^{1,2,3}$} 
\author{Dariusz Kaczorowski$^1$} 
\author{Piotr Wiśniewski$^1$,*}
\affiliation{$^1$Institute of Low Temperature and Structure Research, Polish Academy of Sciences, P.Nr 1410, 50-950 Wrocław, Poland,\\
$^2$Division of Materials Science and Engineering, Ames Laboratory, Ames, Iowa 50011, USA\\
$^3$Department of Physics and Astronomy, Iowa State University, Ames, Iowa 50011, USA}
%
\date{\today}
\begin{abstract}
Monobismuthides of lutetium and yttrium are shown as new representatives of materials which exhibit extreme magnetoresistance and magnetic-field-induced resistivity plateaus. 
At low temperatures and in magnetic fields of 9\,T the magnetoresistance attains orders of magnitude of $10^4\,\%$ and $10^3\,\%$, on YBi and LuBi, respectively.  
Our thorough examination of electron transport properties of both compounds show that observed features are the consequence of nearly perfect carrier compensation rather than of possible nontrivial topology of electronic states. 
The field-induced plateau of electrical resistivity can be explained with Kohler scaling. 
An anisotropic multiband model of electronic transport describes very well the magnetic field dependence of electrical resistivity and Hall resistivity. 
Data obtained from the Shubnikov--de Haas oscillations analysis also confirm that the Fermi surface of each compound contains almost equal amounts of holes and electrons. 
First-principle calculations of electronic band structure are in a very good agreement with the experimental data.\\\\
*~Corresponding author:  p.wisniewski@int.pan.wroc.pl    
\end{abstract}
\maketitle
Materials with extremely magnetic-field-dependent resistivity attract massive attention because of their possible applications in sensors and spintronic devices. 
Rare-earth-metal monopnictides with the NaCl-type crystal structure form a group of materials that possess relevant extraordinary properties. 
The very first observation of extreme magnetoresistance (XMR) in lanthanum monopnictides has been reported by Kasuya {\em et al.} in 1996 \hspace{5pt}\cite{Kasuya1996}.
Two decades later it has been proposed that lanthanum monopnictides could be topologically nontrivial materials \hspace{5pt}\cite{Zeng2015} and magnetotransport properties of LaSb and LaBi have been found to resemble those of topological semimetals \hspace{5pt}\cite{Tafti2016,Tafti2016a}.
It was the starting point of an intensive revival of interest in rare-earth-metal monopnictides.  
Up to date the question of the nontrivial topology of their electronic structures  has remained open. 
Reports on the angle-resolved photoemission spectroscopy (ARPES) investigations of rare-earth-metal monopnictides differ in their conclusions. 
Some describe these materials as having Dirac-like features in their electronic structure \hspace{5pt}\cite{Niu2016,Wu2016b,Nayak2017,Lou2016,Neupane2016a}, others show that  nontrivial topology is absent \hspace{5pt}\cite{Zeng2016,Wu2017b,Oinuma2017,He2016a}.

Non-saturating (in magnetic field) XMR has earlier been reported for Dirac semimetals Cd$_3$As$_2$ and \mbox{ZrSiS}, and Weyl semimetals NbP and TaAs \hspace{5pt}\cite{Liang2014,Shekhar2015b,Huang2015,Lv2016b}.
However, their XMR could be often understood without invoking nontrivial topology. In non-magnetic materials, charge carrier compensation \hspace{5pt}\cite{Mun2012}, field-induced metal-insulator transition \hspace{5pt}\cite{Li2016i} (all unrelated to nontrivial topology), or field-induced lifting of topological protection from backscattering \hspace{5pt}\cite{Liang2014} could be responsible for XMR. 

This work on YBi and LuBi is a continuation of our previous investigations of NaCl-type monoantimonides with high magnetoresistance \hspace{5pt}\cite{Pavlosiuk2016f,Pavlosiuk2017a}.
These two compounds have been barely studied previously. 
The first report on YBi crystal structure appeared in Ref.~\onlinecite{Iandelli1961}, and then binary phase diagrams Y-Bi and Lu-Bi, including YBi and LuBi, have been determined \hspace{5pt}\cite{Schmidt1969,Yoshihara1975,Abdusalyamova1996}.
Several theoretical papers on lutetium monopnictides and YBi also appeared in the past few years \hspace{5pt}\cite{Gupta2013,Ameri2016,Acharya2017}.
There was hitherto no information about magnetotransport properties of yttrium and lutetium monobismuthides. 
Here we report on electronic transport properties of high-quality single crystals of YBi and LuBi studied in magnetic fields up to 9\,T. 
Experimental data are compared with results of electronic structure calculations.    
    
We grew high-quality single crystals from Bi flux with the starting atomic composition $RE$:Bi of 1:19 ($RE$= Y or Lu). They had shapes of cubes with the dimensions up to $4\!\times\!4\!\times\!4$\,mm$^3$. Microanalysis of the crystals with a scanning electron microscope equipped with energy-dispersive X-ray spectrometer (FEI SEM with an EDAX Genesis XM4 spectrometer) yielded equiatomic chemical composition of both compounds. 
Electrical resistivity and Hall effect measurements were carried out in a temperature range from 2 to 300\,K and in applied magnetic fields up to 9\,T on a Quantum Design PPMS platform. Standard four-probe method was used for all measurements.
Bar-shaped specimens with all edges along $\langle100\rangle$ crystallographic directions were cut from single crystals and then polished. Electrical contacts were made from $50\,\mu$m-thick silver wires attached to the samples by spot welding and strengthened with silver epoxy. 
The electric current was always flowing along the $[100]$ crystallographic direction and the magnetic field was applied along the $[001]$ crystallographic direction.
\begin{figure*}
	\includegraphics[width=16cm]{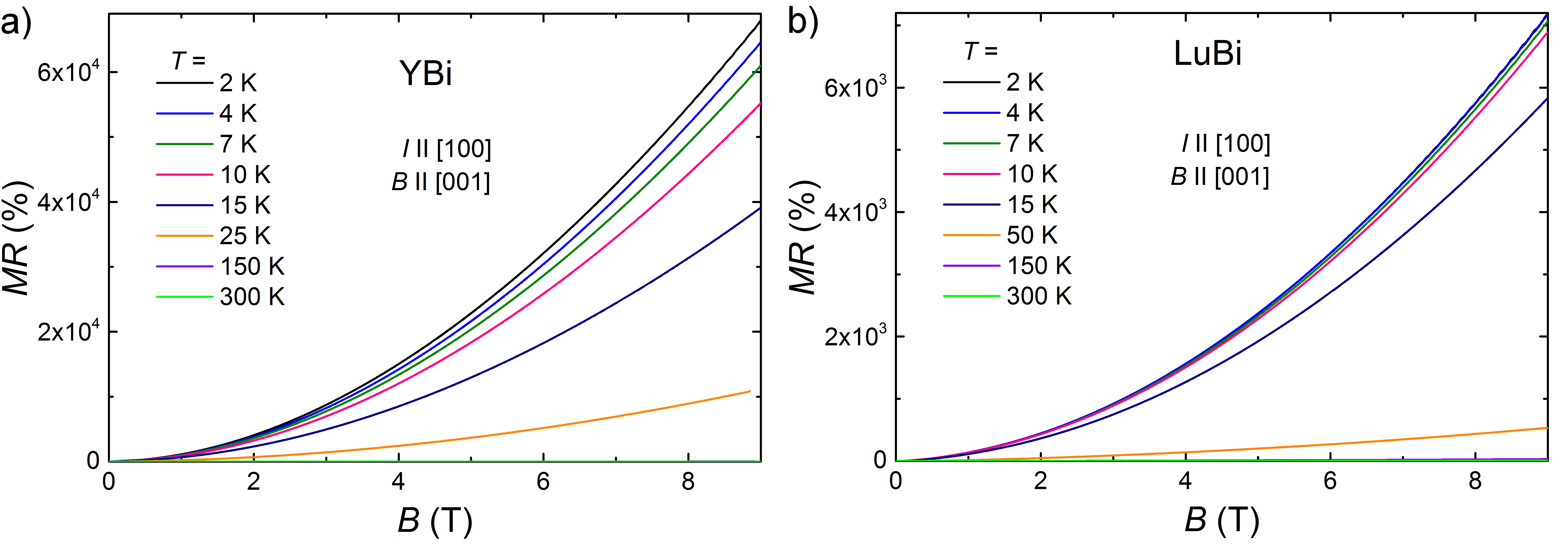}
	\caption{Magnetoresistance isotherms of YBi (a) and LuBi (b) measured in magnetic field applied along [001] direction, transverse to electrical current.  
		\label{MR_all}}
\end{figure*}
\newpage
Electronic structure calculations were carried out using both the {\sc wien2k} 
code \hspace{5pt}\cite{wien2k} with the full-potential linearized augmented plane wave (FLAPW) method, and the full-potential Korringa-Kohn-Rostoker (KKR) 
Green's function method \hspace{5pt}\cite{Ebert2011}. The exchange and correlation effects were treated using the generalized gradient approximation (GGA) \hspace{5pt}\cite{Perdew1996}. 
Spin-orbit coupling was included as a second variational step, using scalar-relativistic eigenfunctions as the basis, after the initial calculation was converged 
to self-consistency. The Monkhorst-Pack special $k$-point scheme with 
$44\!\times\!44\!\times\!44$ mesh was used in the first Brillouin zone sampling, and the cutoff parameter ($R_{mt}K_{max}$) was set to 8.
For the Fermi surface, the irreducible Brillouin zone was sampled by 20225 $k$-points to ensure accurate determination of the Fermi level \hspace{5pt}\cite{Kokalj2003}. Shubnikov--de Haas (SdH) frequencies were calculated using the Supercell $k$-space Extremal  Area Finder tool \hspace{5pt}\cite{Rourke2012}.
\subsection*{Magnetoresistance, electrical resistivity and Hall resistivity} 
Figure~\ref{MR_all} shows magnetoresistance, $M\!R\!=\!100\,\%\!\times\![\rho(B)-\rho(B\!=\!0)]/\rho(B\!=\!0)$, of YBi and LuBi as a function of magnetic field, $B$,  measured at several temperatures, $T$, in the range from 2 to 300\,K. 
For both compounds, $M\!R$ has extreme values at low temperatures (for YBi, $M\!R\!=\!6.8\!\times\!10^4\,\%$ and for LuBi, $M\!R\!=\!7.2\!\times\!10^3\,\%$ at $T\!=\!2$\,K in $B\!=\!9$\,T). 
Up to $T\!=\!10$\,K, magnitudes of $M\!R$ change only slightly, a pattern which corresponds to the resistivity plateaus in $\rho(T)$ (see Fig.~\ref{rho(T)_in_fields}). 
We suppose that such big $M\!R$ of our samples could be due to nearly perfect carrier compensation, as it has been reported for other rare-earth-metal monopnictides  \hspace{5pt}\cite{Zeng2016,Kumar2016,Sun2016a,Niu2016,Ghimire2016,Han2017a,Xu2017f, Pavlosiuk2016f,Pavlosiuk2017a}.

The difference between $M\!R$ values of LuBi and YBi seems to reflect the difference in sample quality, rather than difference in electronic structures (see the next subsection).  
On the example of compensated semimetal WTe$_2$ and lanthanum monopnictides, it has been shown that magnitude of $M\!R$ strongly depends on sample quality \hspace{5pt}\cite{Ali2015, Tafti2016a}. On heating above 10\,K, $M\!R$ of both compounds decreases strongly, and at 300\,K reaches 6$\%$ and 7$\%$ (in 9\,T) for LuBi and YBi, respectively.

\begin{figure}[b]
	\includegraphics[width=7cm]{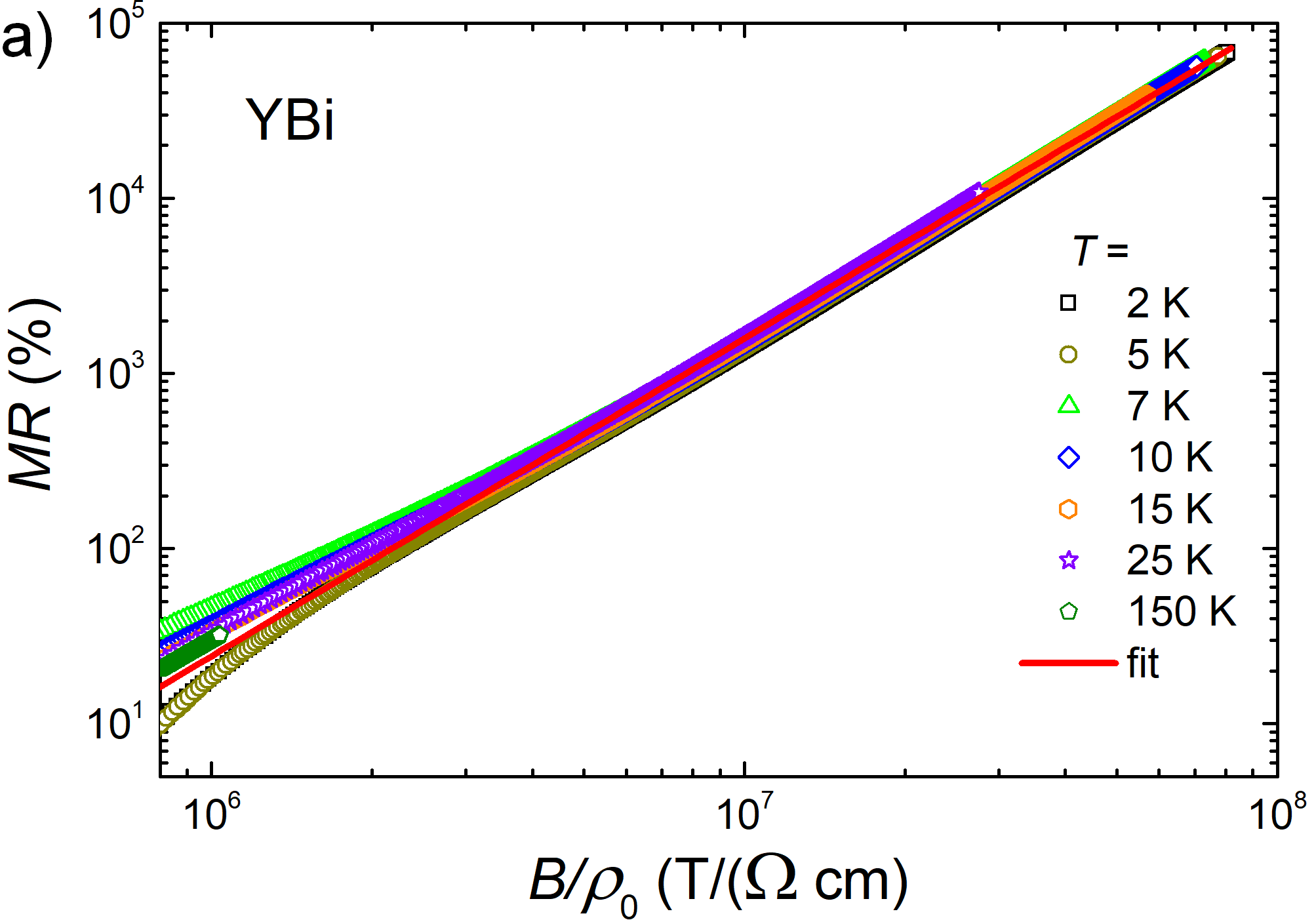}\\
	\includegraphics[width=7cm]{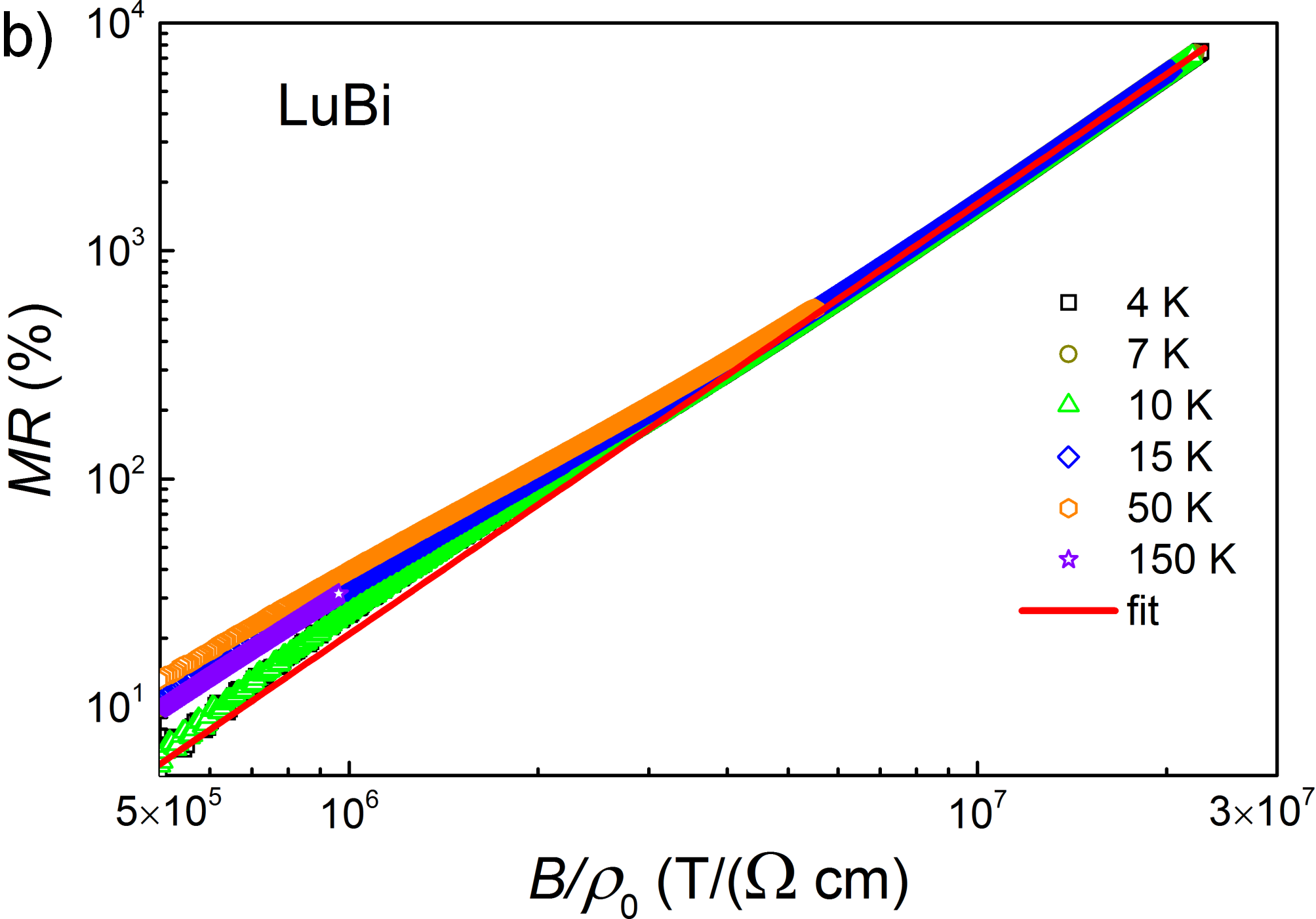}
	\caption{Kohler scaling of transverse magnetoresistance, with $M\!R\propto(B/\rho_0)^{m}$ fitted to the data from temperature range 2--300\,K yielding $m=1.81$ for YBi (a) and $m=1.89$ for LuBi (b).  
		\label{Kohler_scaling}}
\end{figure}

\begin{figure}[h]
	\includegraphics[width=8cm]{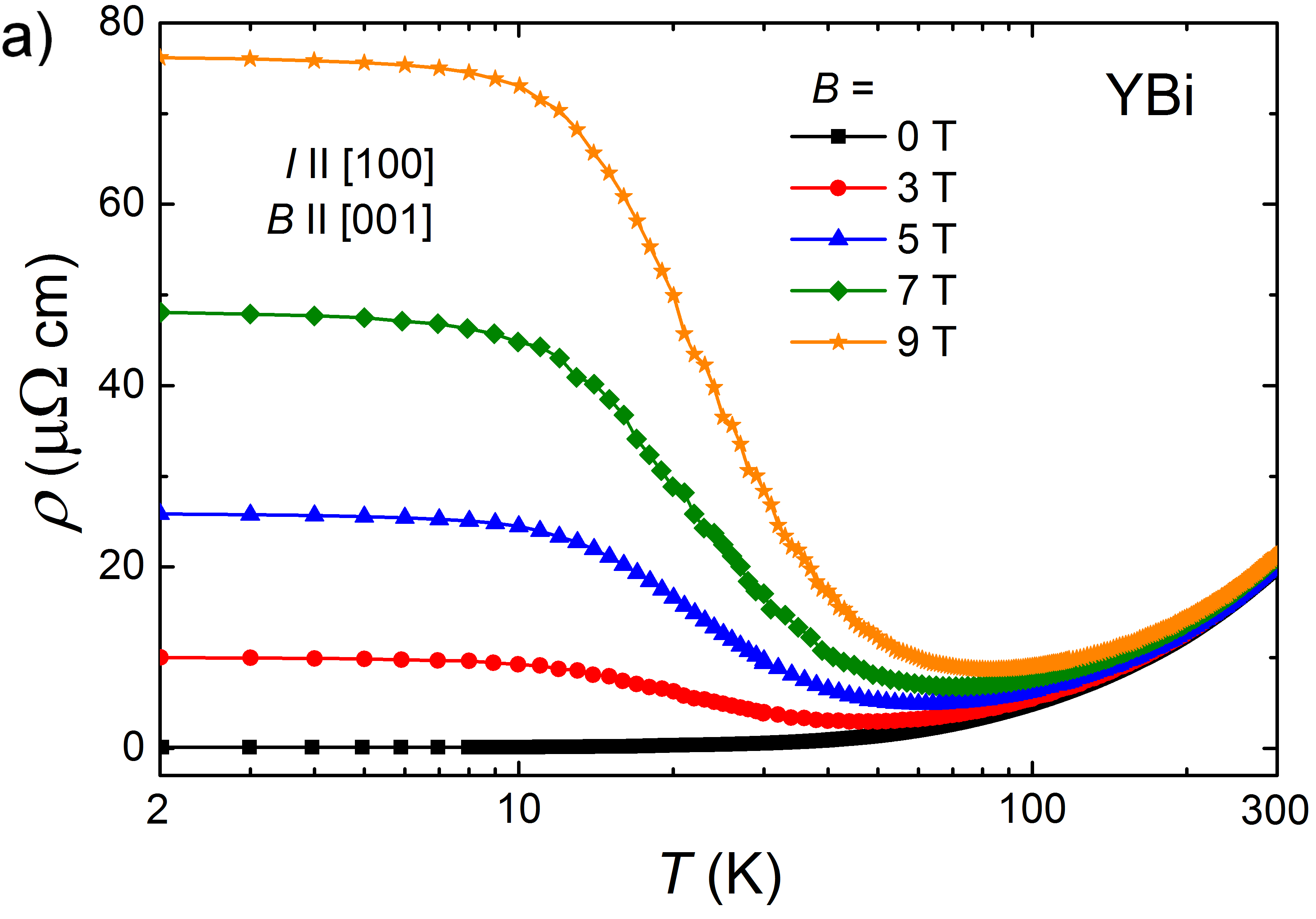}\\
	\includegraphics[width=8cm]{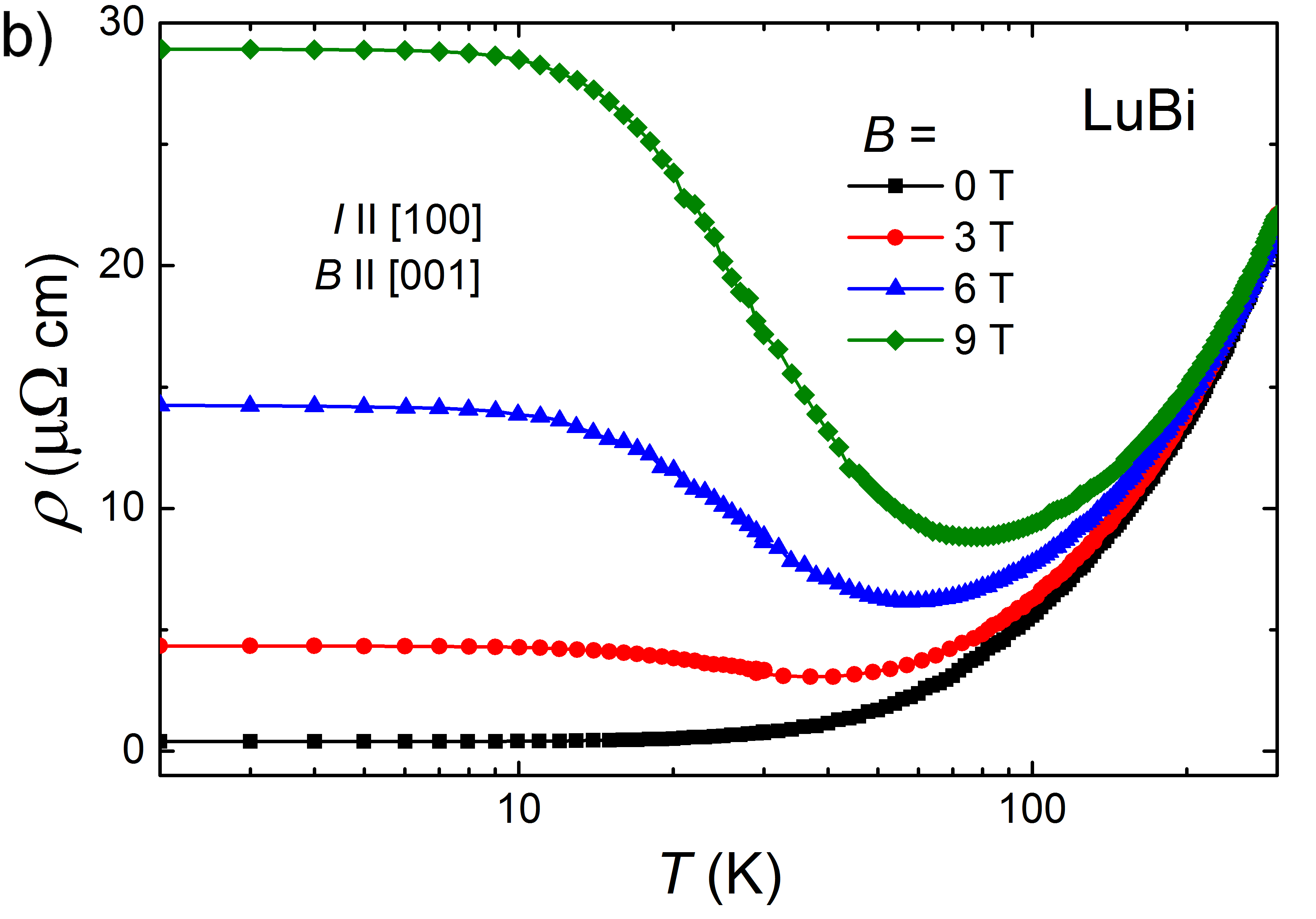}
	\caption{Temperature variations of electrical resistivity of YBi (a) and LuBi (b) in various magnetic fields applied perpendicular to the current direction.  
		\label{rho(T)_in_fields}}
\end{figure}

Figure~\ref{Kohler_scaling} shows the results of Kohler scaling of $M\!R$ for both compounds. 
All $M\!R$ isotherms measured at different temperatures collapse on a single curve. According to the Kohler rule,
\begin{equation}
M\!R\propto(B/\rho_0)^{m},
\label{Kohler_eq}
\end{equation}
where $m$ is a sample-dependent constant that depends on the level of compensation (for perfectly carrier compensated systems $m=2$).
From the fitting of Eq.~\ref{Kohler_eq} (red solid lines in Fig.~\ref{Kohler_scaling}) to experimental data we obtained $m=1.81$ and $m=1.89$ for YBi and LuBi, respectively.

These values of $m$ are larger than previously reported for rare-earth-metal monoantimonides \hspace{5pt}\cite{Pavlosiuk2016f,Pavlosiuk2017a,Han2017a}.
Our $m$ values are close to that determined for LaBi \hspace{5pt}\cite{Sun2016a}, but still smaller than 1.92 reported for WTe$_2$ in Ref.~\onlinecite{Wang2015d}. They show that the carrier compensation in LuBi is slightly better than in YBi.

Figure~\ref{rho(T)_in_fields} presents the results of electrical resistivity, $\rho$, measurements for YBi and LuBi in varying temperature in zero and in finite magnetic fields. 
When $B=0$, both compounds demonstrate metallic behavior of $\rho(T)$, 
$\rho$ gradually decreases with $T$ lowering, from the values 20.0 and $21.6\,\mu\Omega\,\rm{cm}$ at $T=300$\,K to the values 0.1 and $0.4\,\mu\Omega\,\rm{cm}$ at $T=2\,$K for YBi and LuBi, respectively. 
It means that residual resistivity ratios [$\rho(300\,\rm{K})/\rho(2\,\rm{K})$] are quite large and equal to 180 and 55 for YBi and LuBi, respectively. 

Applying a magnetic field drastically changes the $\rho(T)$ behavior. Already in 3\,T, $\rho$ of each compound decreases upon cooling only to certain temperature where it has a minimum. Further decreasing of temperature leads to increase of $\rho$ and its saturation below $T\approx10\,$K. Higher fields increase the values of resistivity in plateau region in accordance with $M\!R\propto B^m$ behavior depicted in Fig.~\ref{MR_all}.
Such magnetic field-induced resistivity plateau is a characteristic feature of topological semimetals \hspace{5pt}\cite{Shekhar2015b,Li2016i,Hosen2017} and has also been observed in several rare-earth-metal monopnictides \hspace{5pt}\cite{Tafti2016a,Ghimire2016,Kumar2016,Pavlosiuk2016f,Pavlosiuk2017a,Wu2017d}.

\begin{figure}[t]
   \includegraphics[width=8cm]{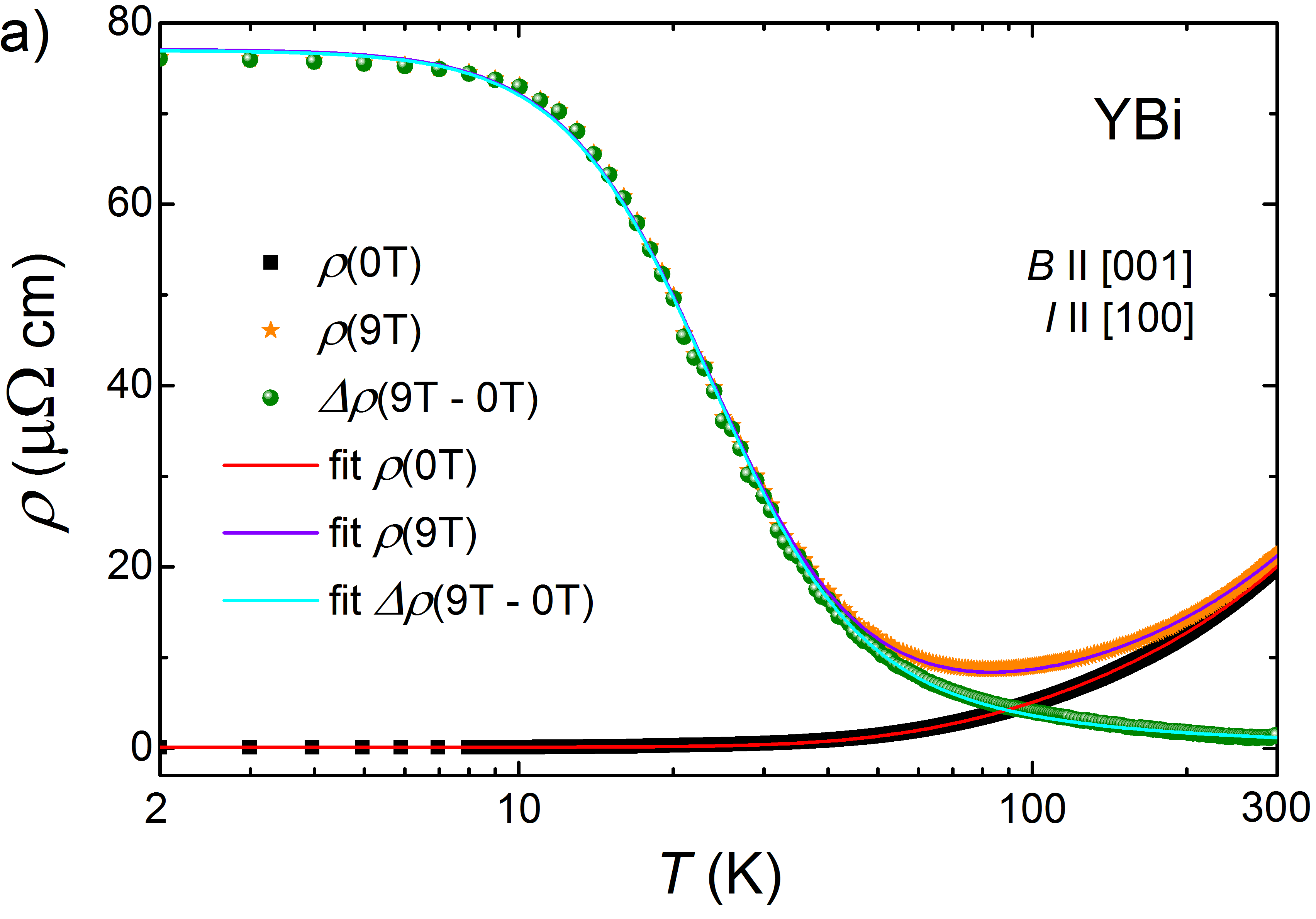}\\
	\includegraphics[width=8cm]{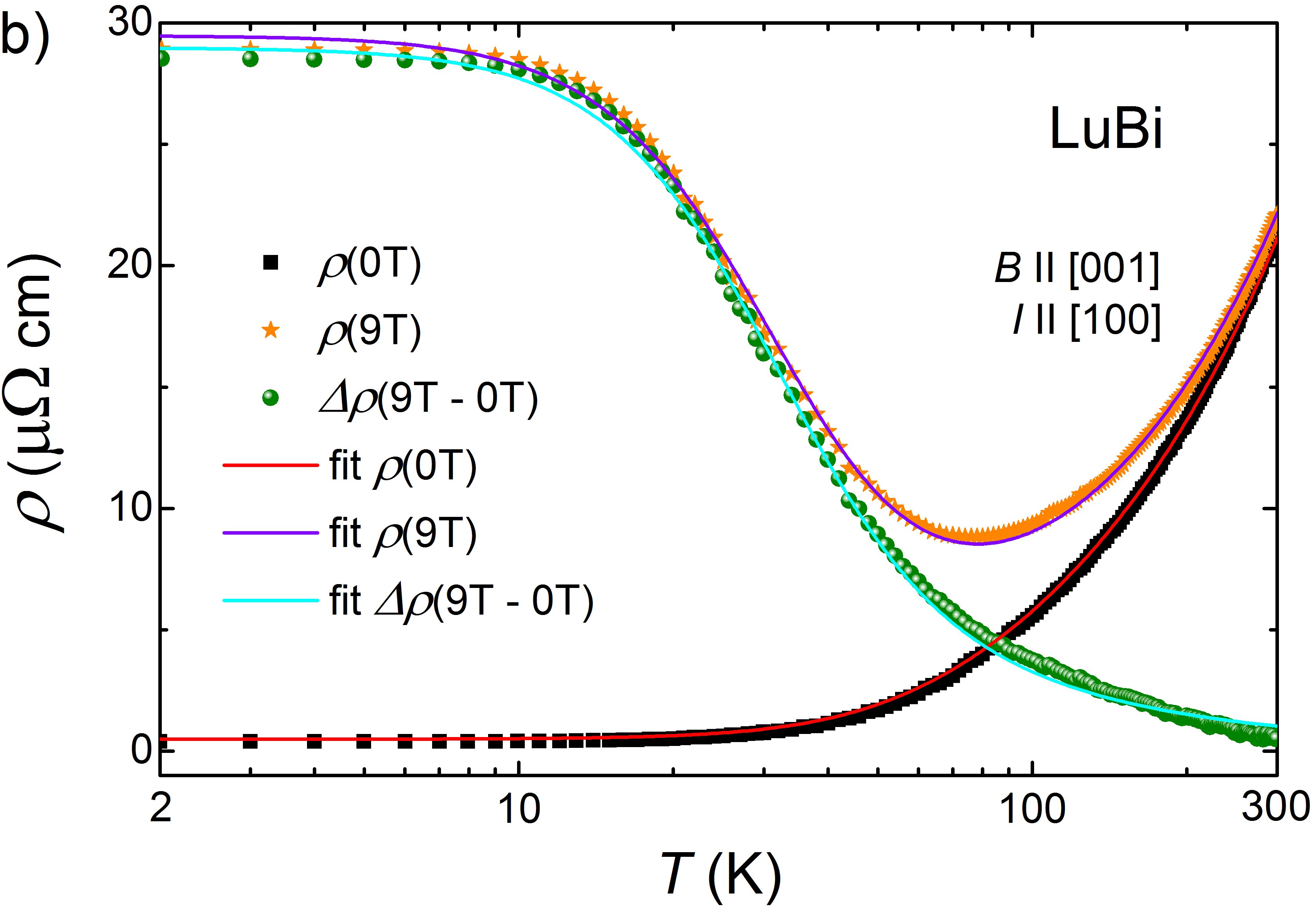}
	\caption{Temperature variations of electrical resistivity measured in magnetic fields of 9 and 0\,T and their difference for YBi (a) and LuBi (b). The solid lines correspond to fits of Eqs.~\ref{BG_eq+Kohler_eq} and \ref{BG_eq}. 
		\label{rho(T)_analiza}}
\end{figure}

The authors of Ref.~\onlinecite{Wang2015d} argued that analogous turn-on behavior of $\rho(T)$ in WTe$_2$ could be understood in the scope of Kohler scaling. We used this approach to describe electrical resistivity of both studied monopnictides (see Fig.~\ref{rho(T)_analiza}). 
Previously, it has also been used by Han {\em et al.} to explain magnetotransport properties of LaSb \hspace{5pt}\cite{Han2017a}. 
According to Wang {\em et al.} \hspace{5pt}\cite{Wang2015d}, $\rho(T)$ measured in magnetic field can be described by the following equation:
\begin{equation}
\rho(T,B)=\rho_0(T,0)+\Delta\rho(T,B),
\label{BG_eq+Kohler_eq}
\end{equation}
where the first term corresponds to the temperature dependence of resistivity in zero magnetic field and the second term describes magnetic-field-induced resistivity. 
Assuming that $\rho_0(T,0)$ can be well approximated with the Bloch-Gr{\"u}neisen law,
\begin{equation}
\rho(T)=\rho_0+A\Bigg(\frac{T}{\Theta_D}\Bigg)^k\int_{0}^{\frac{\Theta_D}{T}} \frac{x^k}{(e^x-1)(1-e^{-x})} dx,
\label{BG_eq}
\end{equation}
and
\begin{equation}
\Delta\rho(T,B)=\gamma B^m/[\rho_0(T,0)]^{m-1},
\label{delta_BG_eq+Kohler_eq}
\end{equation}

we simultaneously fitted $\rho(T)$  in zero field with Eq.~\ref{BG_eq} and $\rho(T)$  measured in $B=9$\,T with Eq.~\ref{BG_eq+Kohler_eq} using shared parameters. 
Fits to these model with $\rho_0$, $A$, $k$, $\Theta_D$, and $\gamma$ as free parameters, and parameter $m$ fixed at its value obtained from Kohler scaling are shown as red and purple solid lines in Fig.~\ref{rho(T)_analiza}.
The obtained parameters for both compounds are rather similar and listed in Table~\ref{B-G_table}. 
\begin{table}[b]
   \begin{ruledtabular}
	\begin{tabular}{l *{6}{c}} 
&$\rho_0~~$&$A$~~~&$\Theta_D$~~~&$k$&$m$&$\gamma$\\ 
&($\mu\Omega\,$cm)~~~&($\mu\Omega\,$cm)~~~&(K)~~~&~~~&~~~&($\Omega\,$cm)$^{m}$ \\\colrule
		YBi& 0.12~~& 42.3~~& 295~~~& 3.08~~& 1.81~~&$3.3\!\times\!10^{-12}$\\
		LuBi& 0.5~~& 34~~~& 307~~~& 2.55~~& 1.89~~&$1.1\!\times\!10^{-12}$\\
	\end{tabular}
\end{ruledtabular}
\caption{Parameters obtained from the fitting of Eqs.~\ref{BG_eq+Kohler_eq} and \ref{BG_eq} to the $\rho(T)$ data, as shown in Fig.~\ref{rho(T)_analiza}.}
	\label{B-G_table}
\end{table}
Values of $k$ are close to that previously determined for LuSb \hspace{5pt}\cite{Pavlosiuk2017a} and several Lu- and La-containing intermetallics \hspace{5pt}\cite{Fournier1993}. 
The Debye temperatures are smaller than $\Theta_D=408$ and 420\,K reported for LuSb and LuAs, respectively \hspace{5pt}\cite{Pavlosiuk2017a,Zogal2014}.

Additionally, we show in the Fig.~\ref{rho(T)_analiza} magnetic field-induced resistivity versus temperature as a green circles. 
This data were obtained by subtraction of data measured in zero magnetic field from those measured in 9\,T. 
Cyan solid lines in Fig.~\ref{rho(T)_analiza} represent Eq.~\ref{delta_BG_eq+Kohler_eq} with parameters yielded by the fitting of Eq.~\ref{BG_eq+Kohler_eq}. 
In order to get more insight in carrier concentration we measured Hall resistivity ($\rho_{xy}$) at the temperature of 2 K, where MR attains its maximum.
The $\rho_{xy}(B)$ plots for both compounds are shown in insets to Figs.~\ref{YBi_MR_analiza}a and \ref{YBi_MR_analiza}b. Their curved shapes indicate multiband character of conductivity. Since $\rho_{xy}<<\rho_{xx}$ for both compounds, in further analysis we use Hall conductivity $\sigma_{xy}$ calculated using Eq.~\ref{sig_conversion_rho}.
\subsection*{Electronic structure calculations and Shubnikov--de Haas effect}
\begin{figure}[h]
	\includegraphics[width=7.5cm]{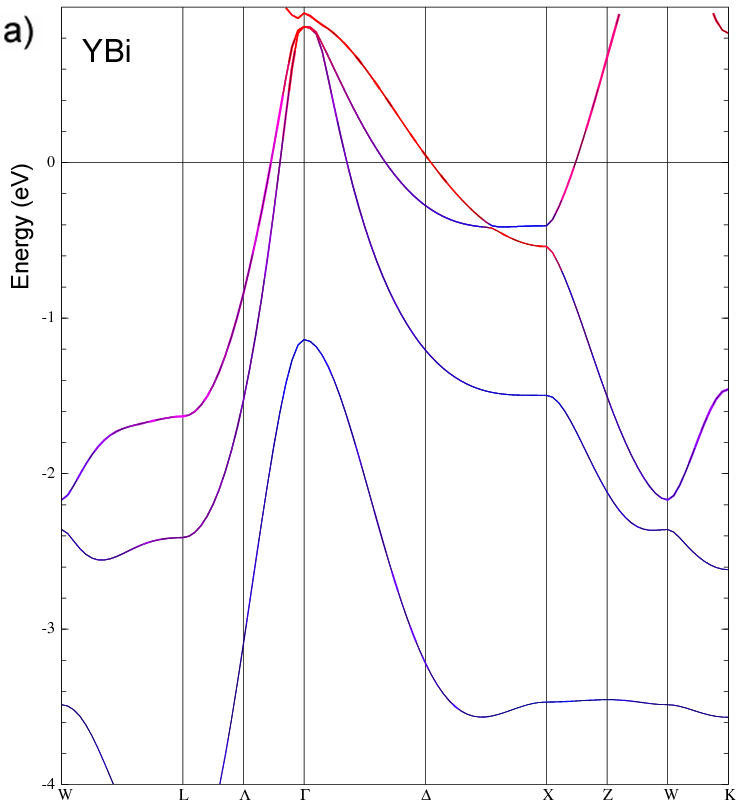}\\
	\includegraphics[width=7.5cm]{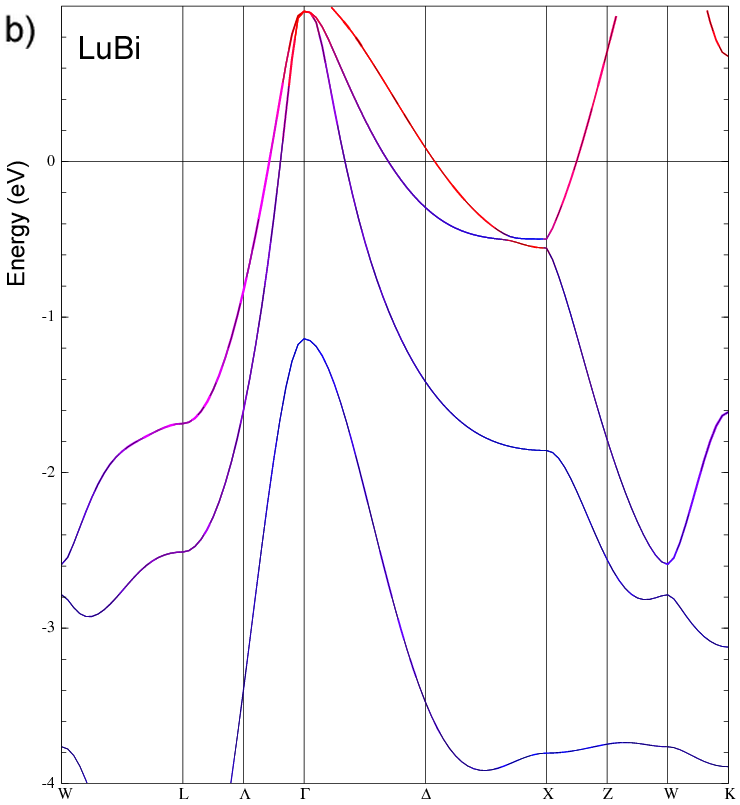}
	\caption{Electronic band structure of YBi (a) and LuBi (b). Horizontal line marks the Fermi level. Red and blue colors denote contributions from $d$-electrons of Y or Lu, and $p$-electrons of Bi, respectively.
		\label{el_struct}}
\end{figure}
Figure~\ref{el_struct} presents calculated bulk electronic band structures of YBi and LuBi. 
The results of our calculations are consistent with scalar-relativistic data 
obtained for YBi \hspace{5pt}\cite{Acharya2017}.  
Both compounds have very similar electronic structures. 
Due to spin-orbit interaction three-fold degeneracy of Bi-$6p$ states is modified at the $\Gamma$ point, i.e. one of the $p$-bands dips deeply below $E_{\rm F}$, whereas two  other $p$-bands remains degenerated and stay above $E_{\rm F}$. Furthermore, the two-fold degeneracy of these two bands is gradually lifted  along $\Gamma\!-\!$L and $\Gamma\!-\!$X lines, and they become well separated at points L and X. 
The corresponding shifts of $p$-bands are noticeably smaller in analogous monoantimonides \hspace{5pt}\cite{Pavlosiuk2016f,Pavlosiuk2017a}, and eventually become just-noticeable in arsenides (data not shown), reflecting decreasing spin-orbit coupling strength.

\begin{figure}[h]
	\includegraphics[width=6cm]{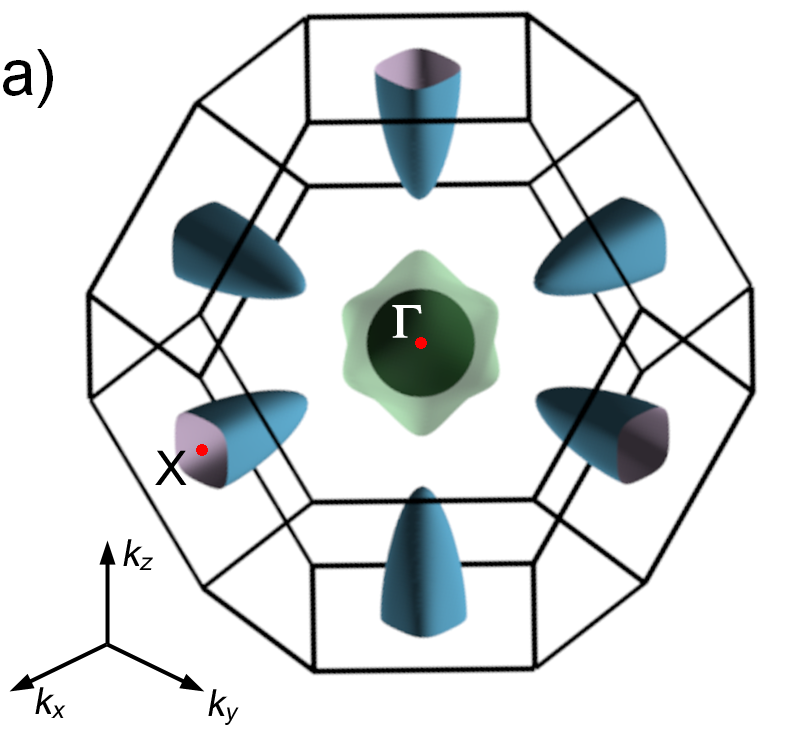}
   \includegraphics[width=5.5cm]{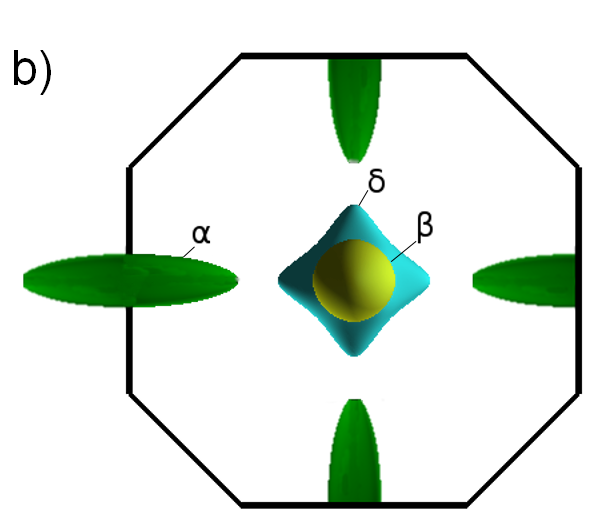}
	\caption{(a) Fermi surface of YBi and LuBi. It consists of a triplicate electron pocket $\alpha$ and two hole pockets $\delta$ and $\beta$. (b) Projection of the Fermi surface on the (001) plane. Proportions between the Brillouin zone and Fermi pockets sizes were not preserved.        
		\label{el_struct_2}}
\end{figure}

Fermi level crosses two hole-like bands near the $\Gamma$ point of Brillouin zone and one electron-like band around the X point. 
Besides, at $\approx0.5$\,eV below the Fermi level, there is a tiny gap between the bands and the band inversion occurs. This is where the Dirac cones potentially may form. 
Analogous gaps have previously been reported in lanthanum monopnictides \hspace{5pt}\cite{Zeng2015,Guo2016d} and YSb \hspace{5pt}\cite{Pavlosiuk2016f}, calculated using the GGA with Perdew-Burke-Ernzerhof exchange-correlation potential. 
Our electronic structure calculations reveals also some $d\!-\!p\,-$mixed orbital texture near the X point of the Brillouin zone (visualized with red and blue colors in Fig.~\ref{el_struct}). 
This finding resembles those for PtSn$_4$, NbSb$_2$, LaBi and WTe$_2$ \hspace{5pt}\cite{Jiang2015,Tafti2016a}. 
The Fermi surface is very similar in both compounds and thus schematically depicted, together with its projection on $(001)$ plane, in a common Fig.~\ref{el_struct_2}.  
It consists of a triplicate electron pocket centered at the X points (denoted as $\alpha$) and two hole pockets ($\beta$ and $\delta$) nested in the center of the Brillouin zone. 
The calculations of electronic structure brought also the carrier concentrations, $n_p^{calc}$, cyclotron frequencies for maximal cross-sections of Fermi pockets by planes perpendicular to $[001]$ direction, $f_p^{calc}$, and corresponding cyclotron masses, $m_p^{*calc}$. Their values are collected in Table~\ref{FFT_TABLE}. 
Comparing the ratios of the concentrations of electrons and holes $n_{\alpha}^{calc}/(n_{\beta}^{calc}+n_{\delta}^{calc})$, being 1.003 in YBi and 1.002 in LuBi, suggests that carrier compensation is nearly perfect in both compounds. 

\begin{figure}[h]
	\includegraphics[width=8cm]{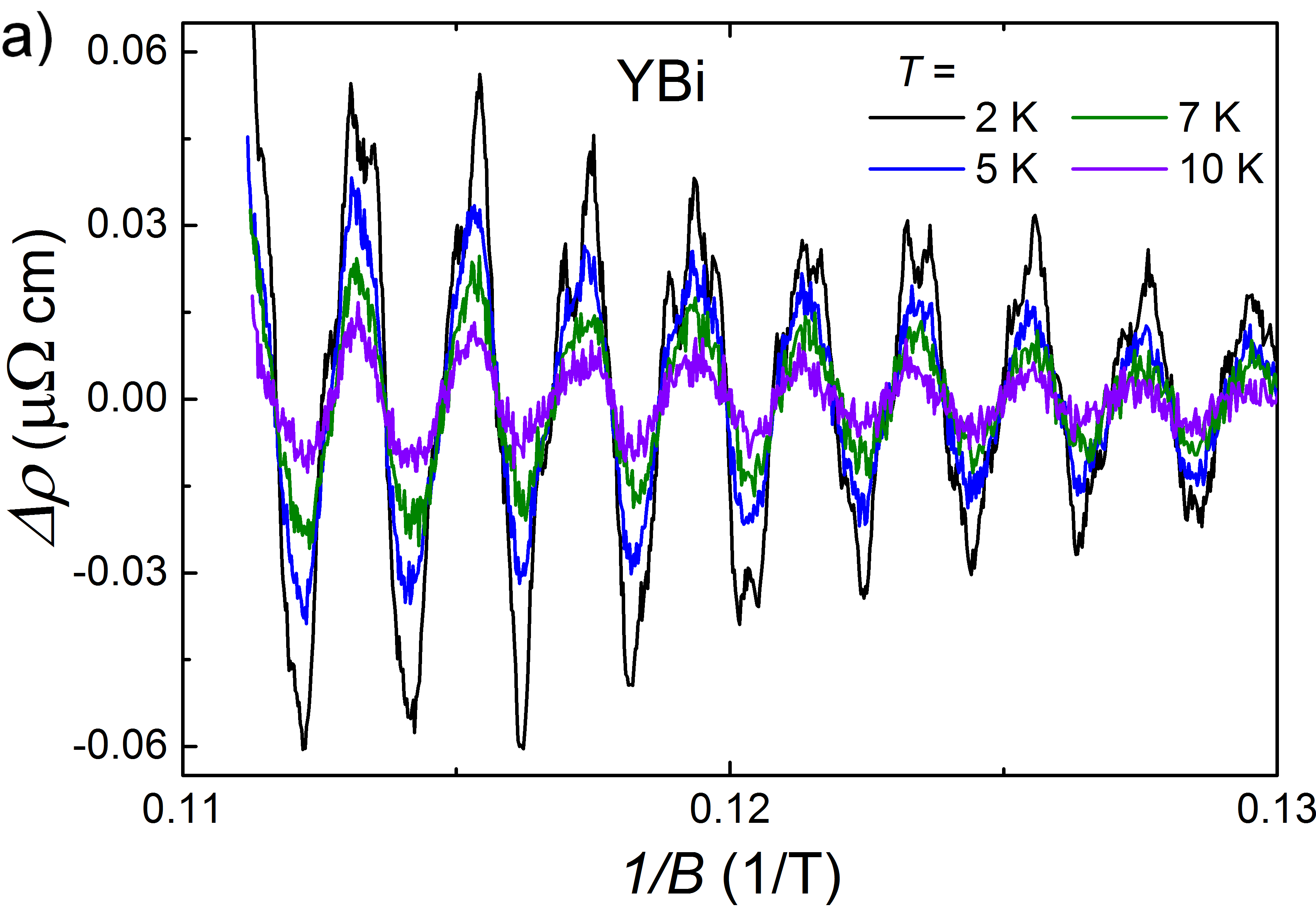}
   \includegraphics[width=8cm]{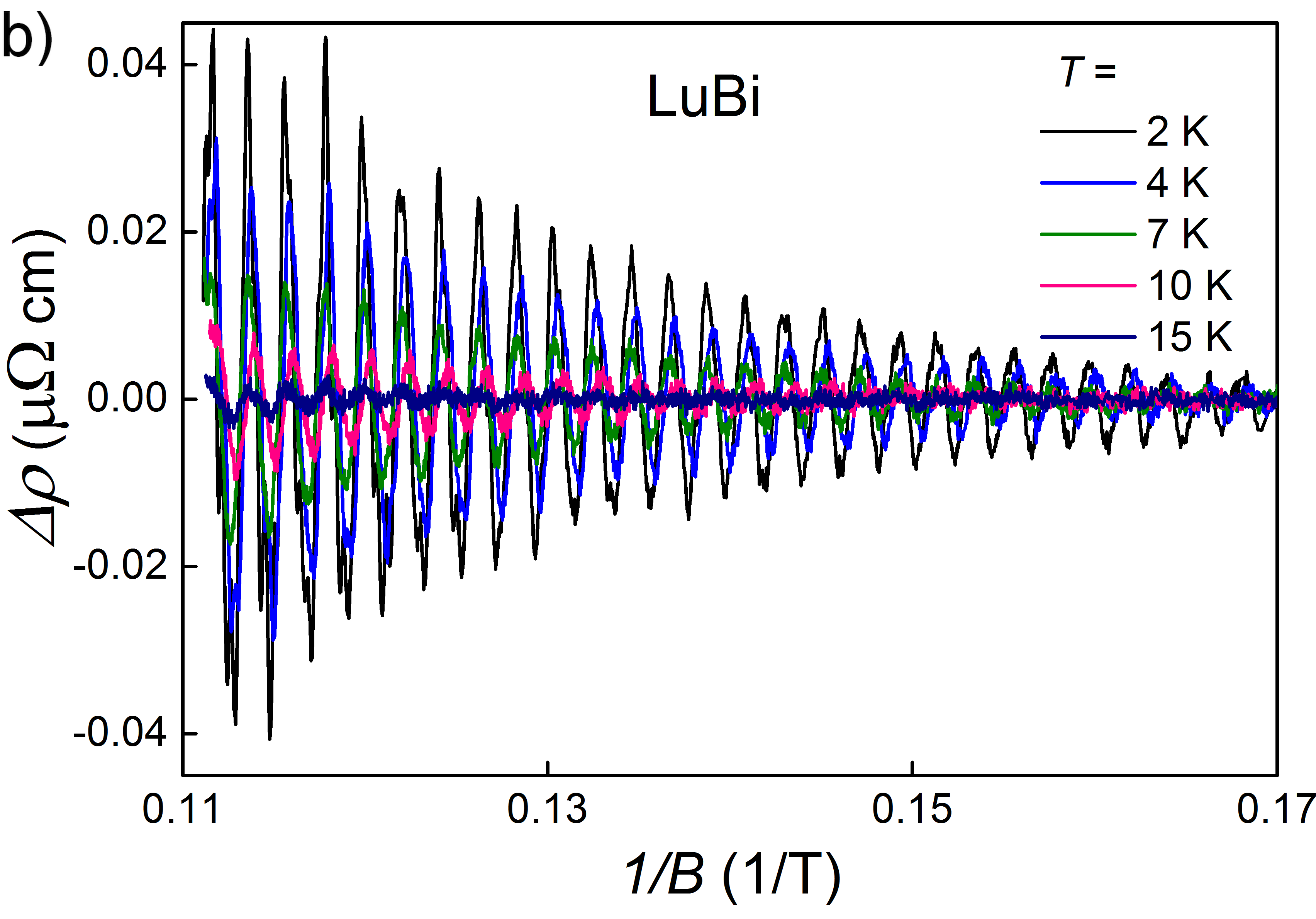}
	\caption{Oscillating part of electrical resistivity as a function of inverted magnetic field for YBi (a) and LuBi (b), measured at several different temperatures.  
		\label{SdH_effect}}
\end{figure}
\begin{figure*}
	\includegraphics[width=16cm]{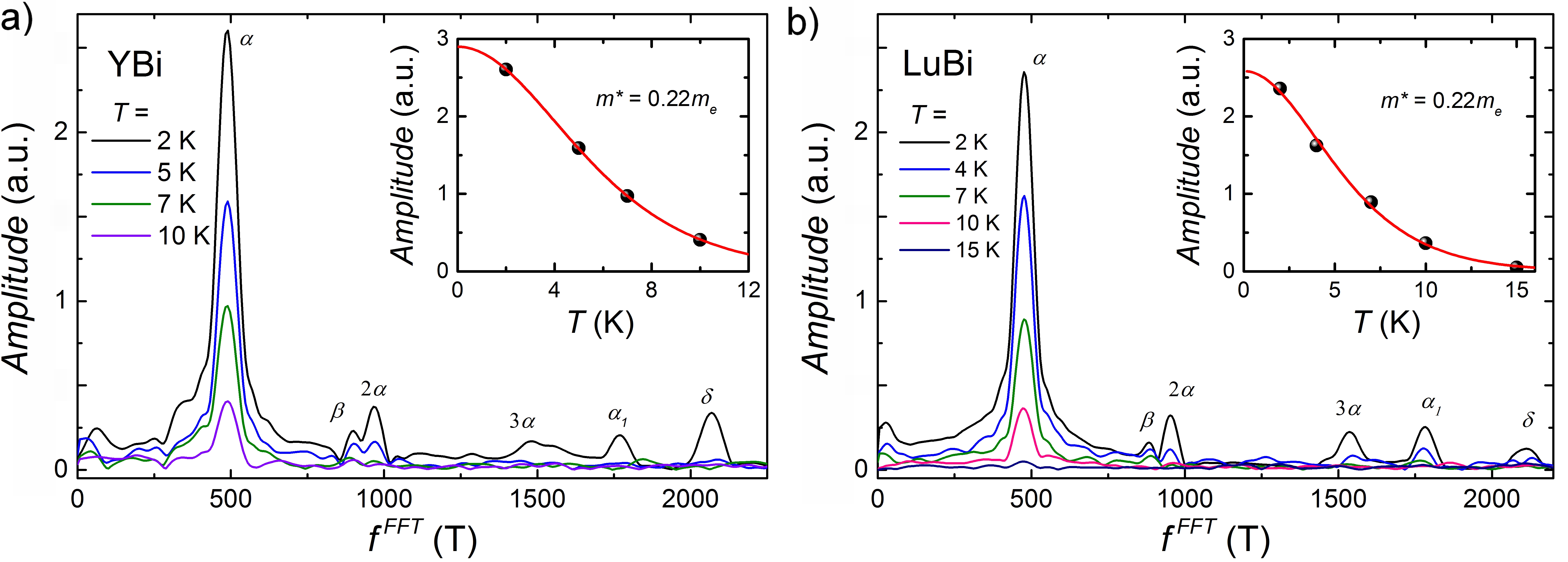}
	\caption{Fast Fourier transform analysis of oscillating part of electrical resistivity of YBi (a) and LuBi (b). Insets: temperature dependence of the amplitude of the highest peak in the FFT spectra. Red solid line represents fits of Eq.~\ref{LK_eq} to the experimental data.  
		\label{FFT}}
\end{figure*}

Good quality of our samples allowed us to observe quantum oscillations of electrical resistivity in magnetic field, i.e., the Shubnikov--de Haas (SdH) effect. 
The subtraction of the third-order polynomial background from the $\rho(1/B)$ data resulted in experimental curves presented in Fig.~\ref{SdH_effect}. 
Strong SdH oscillations were clearly observed at temperatures up to at least 10 and 15\,K for YBi and LuBi, respectively. 
The shape of $\Delta\rho(1/B)$ suggests multifrequency character of the oscillations.  
Indeed, their fast Fourier transform (FFT) analysis shows for each of two compounds, six pronounced maxima (see Fig.~\ref{FFT}). Corresponding SdH frequencies $f_p^{FFT}$ are listed in Table~\ref{FFT_TABLE}. 
These, denoted with $f_{\alpha}^{FFT}$ and $f_{\alpha'}^{FFT}$, we ascribe to the electrons on orbits being maximal cross sections of ellipsoid-like Fermi pocket $\alpha$, perpendicular to its long and short axis, respectively. $f_{2\alpha}^{FFT}$ and $f_{3\alpha}^{FFT}$ are the second and the third harmonics of $f_{\alpha}^{FFT}$.
Frequencies $f_{\beta}^{FFT}$ and $f_{\delta}^{FFT}$ are due to the hole pockets.
We obtained very similar FFT spectra, matching very well the results of our electron structure calculations, for both compounds. 
According to the Onsager relation: $f_{SdH}=hS/e$, where $S$ is the area of Fermi surface cross section \hspace{5pt}\cite{Shoenberg1984}. 
Assuming perfect ellipsoidal shape of the $\alpha$ sub-pockets and the spherical one of pocket $\beta$, we calculated the Fermi wave vectors and than carrier concentrations using the formula $n_p=V_{F,p}/(4\pi^3)$, where $V_{F,p}$ is the volume of Fermi pocket $p$. The $n_e/n_h$ ratios resulting from analysis of SdH oscillations are 0.97 and 0.95 for YBi and LuBi, respectively. This shows that the electron-hole compensation is very close to perfect in both compounds, as hinted above by Kohler scaling and DFT calculations. 

Effective masses ($m^*$) of the carriers of $\alpha$ Fermi pocket were calculated from the temperature dependence of FFT amplitude, $R_\alpha$, at $f_{\alpha}^{FFT}$ frequency, obtained from the field window 7--9\,T, using the following relation \hspace{5pt}\cite{Shoenberg1984}:  
\begin{equation}
R_\alpha(T)\propto(\lambda m^*T/B_{eff})/\sinh(\lambda m^*T/B_{eff}),
\label{LK_eq}
\end{equation}
with $B_{eff} = 7.875$\,T being the the reciprocal of average inverse field from the  window where FFT was performed: $B_{eff}=2/(1/B_1 + 1/B_2)$ (with $B_1=7$\,T and $B_2=9$\,T), and the constant $\lambda=2\pi^2k_Bm_0/e\hbar\,(\approx14.7\,$T/K), we obtained $m^*=0.22\,m_0$ for both compounds. 
This value of effective mass is close to those reported previously for other rare-earth-metal monopnictides \hspace{5pt}\cite{Wu2017d,Pavlosiuk2016f,Pavlosiuk2017a,Kumar2016, Tafti2016,Wakeham2016} and also to effective masses $m_{\alpha}^{*calc}=0.24\,m_0$ and $0.29\,m_0$, obtained from our electronic structure calculations for YBi and LuBi, respectively. 

Observing good agreement of SdH analysis, the calculations and multiband fitting of magnetotransport (described in next section), all revealing or taking into account strong anisotropy of electron pocket, we decided not to pursue angle-dependent SdH measurements as we expect that they would yield results very similar to those presented in other papers on similar monopnictides \hspace{5pt}\cite{Kumar2016,Pavlosiuk2016f, Pavlosiuk2017a,Yang2017c}.
\begin{table*}[h]
\begin{ruledtabular}
	\begin{tabular}{*{9}{l}r}
{Compound}&&~~~~~~~~~~~~~~{$p$\,=}~~&$\alpha$&$\alpha'$&$2\alpha$&$3\alpha$&$\beta$~~~~&$~~~~\delta$&$n_e/n_h$\\
 YBi& $f^{FFT}_p$&(T)&490~~~~~~~&1766~~~~~~~& 966~~~~~~~&1478~~~~~~~&900~~~~~~~&2069\\
 & $k_F$&(Å$^{-1}$)&0.122~~~&0.439~~~&~~~-~~~&~~~-~~~&0.165~~~&0.251~~~\\
 & $n_{p}$&(10$^{20}$cm$^{-3}$)&\multicolumn{2}{c}{6.63~~~~}&~~~-~~~&~~~-~~~&1.53&5.33&0.97\\
 & $f^{calc}_p$&(T)&544~~~&1853~~~&~~~-~~~&~~~-~~~&1018~~~&2492~~~\\
 & $n^{calc}_p$&(10$^{20}$cm$^{-3}$)&\multicolumn{2}{c}{7.52~~~~}&~~~-~~~&~~~-~~~&1.80~~~& 5.70& 1.003 \\
 & $m_p^{*calc}$&($m_0$)&0.24&0.60&~~~-~~~&~~~-~~~&0.20&0.61\\\hline
 LuBi& $f^{FFT}_p$&(T)&477~~~&1784~~~&953~~~&1535~~~&884~~~&2112~~~\\
 &$k_F$&(Å$^{-1}$)& 0.120 & 0.451 &~~~-~~~&~~~-~~~& 0.164& 0.253 \\
 & $n_{p}$&(10$^{20}$cm$^{-3}$)&\multicolumn{2}{c}{6.61~~~~}&~~~-~~~&~~~-~~~&1.49&5.50&0.95\\
 & $f^{calc}_p$&(T)&680~~~&1868~~~&~~~-~~~&~~~-~~~&980~~~&2738~~~\\
 & $n^{calc}_p$&(10$^{20}$cm$^{-3}$)&\multicolumn{2}{c}{8.39~~~~}&~~~-~~~&~~~-~~~&1.73&6.64&1.002\\
 & $m_p^{*calc}$&($m_0$)&0.29&0.56&~~~-~~~&~~~-~~~&0.18&0.59\\
	\end{tabular}
\end{ruledtabular}
	\caption{Parameters obtained from analysis of SdH oscillations measured at $T=2$\,K and from electronic band structure calculations. 
	\label{FFT_TABLE}}
\end{table*}
\subsection*{Multiband model of magnetotransport}
After establishing the presence of three distinct Fermi pockets, we proceeded to analyze how their form determines the field dependence of transverse magnetoresistivity, $\rho_{xx}$, and Hall resistivity, $\rho_{xy}$. 

Cubic crystal symmetry of YBi and LuBi allows us to define components of conductivity tensor as follows:
\begin{equation}
\begin{matrix*}[r]
\sigma_{xx}=&\rho_{xx}/[(\rho_{xx})^2+(\rho_{xy})^2]\\
\sigma_{xy}=&-\rho_{xy}/[(\rho_{xx})^2+(\rho_{xy})^2]. 
\end{matrix*}
\label{sig_conversion_rho}
\end{equation} 

In semiclassical Drude model, conductivities of individual electron and hole pockets (indexed with $p$) are summed up to obtain total transverse and longitudinal components of conductivity tensor as follows:
\begin{equation}
\begin{matrix*}[l]
\sigma_{xx}= \sum_p e\,n_p\mu_p/[1+(\mu_pB)^2]\\\\
\sigma_{xy}= \sum_p e\,n_p\mu_p^2B/[1+(\mu_pB)^2].
\end{matrix*}
\label{sig_tensor_components}
\end{equation} 
Following the idea of Xu {\em et al.} \hspace{5pt}\cite{Xu2017f} and stressing the inadequacy of an isotropic multiband model for the transport properties of a system with anisotropic Fermi pockets, we used the same analysis as those authors, namely an anisotropic three-band model, taking into account pronounced anisotropy of the electron band $\alpha$ by using separate conductivities for pockets elongated parallel and transverse to the current direction, distinguished by two mobilities $\mu_\parallel$ and $\mu_\bot$. 

Since in the case of LaBi several authors used the effective two-band model, neglecting the anisotropy of electron pocket \hspace{5pt}\cite{Tafti2016,Kumar2016,Sun2016a}, we also tested that  model for YBi and LuBi. However, the fittings with the three-band model were clearly better (see the Supplemental Material \hspace{5pt}\cite{SupplMat}). 
\begin{figure*}
	\includegraphics[width=16cm]{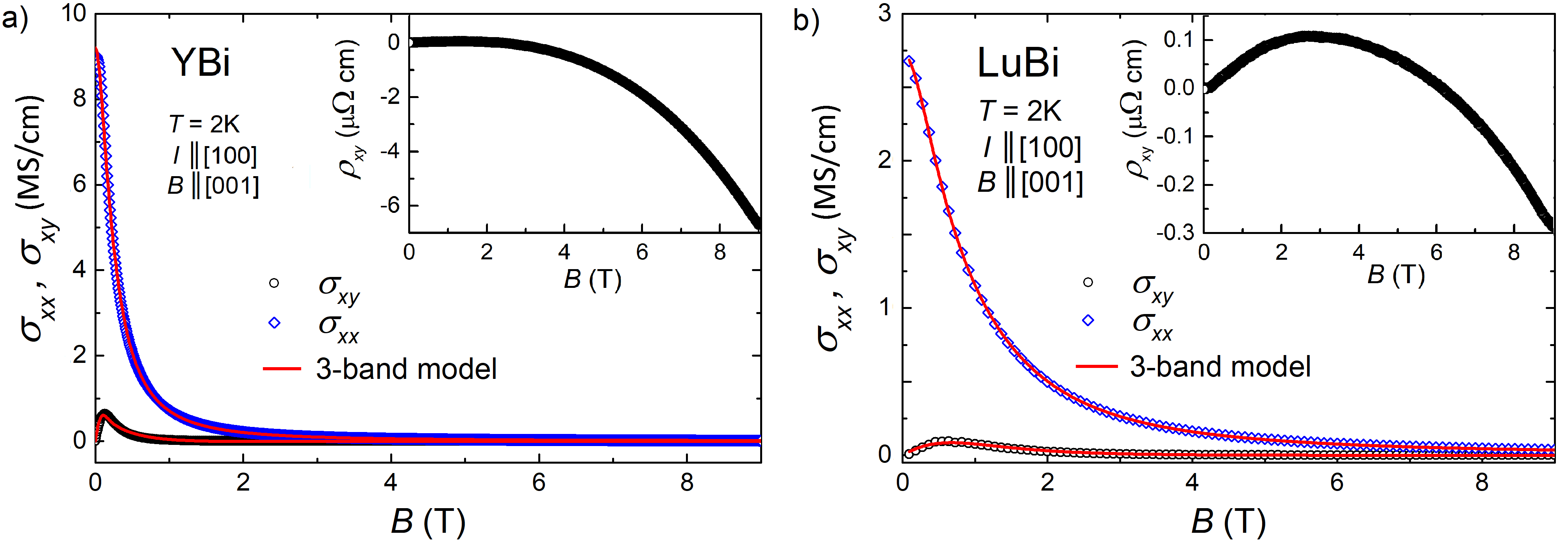}
	\caption{Electrical conductivity and Hall conductivity versus magnetic field measured at $T=2$\,K of (a) YBi and (b) LuBi. Red lines correspond to the fits with Eqs.~\ref{sig_tensor_components}  
		\label{YBi_MR_analiza}}
\end{figure*}

We fitted simultaneously both $\sigma_{xx}$ and $\sigma_{xy}$ of Eq.~\ref{sig_tensor_components} to $\sigma_{xx}(B)$ and $\sigma_{xy}(B)$ data recorded at $T=2$\,K, with shared parameters [using as $\rho_{xx}(B)$ the data shown in Fig.~\ref{MR_all} plots of $M\!R$ for 2\,K].  
Resulting $n_\alpha$, $n_\beta$, and $\kappa(\equiv\mu_{\bot}/\mu_{\parallel}$), together with $\mu_{\bot}$, $\mu_\beta$, $n_\delta$, and $\mu_\delta$ obtained from the fitting of Eq.~\ref{sig_tensor_components} are listed in Table~\ref{MR_ANALIZA_TABLE}. 
\begin{table*}[b]
	\begin{ruledtabular}
	\begin{tabular}{l *{8}{c}} 
{Compound~~~}&$n_\alpha$&$\mu_\bot$&$n_\beta$&$\mu_\beta$&$n_\delta$&$\mu_\delta$&$\kappa$&$n_e/n_h$ \\
&$(\rm{cm}^{-3})$&$(\rm{m^2V^{-1}s^{-1}})$&$(\rm{cm}^{-3})$&$(\rm{m^2V^{-1}s^{-1}})$&$(\rm{cm}^{-3})$&$(\rm{m^2V^{-1}s^{-1}})$ &$~~~~~~$ \\\hline
YBi&$6.88\!\times\!10^{20}$& 6.92&$2.37\!\times\!10^{20}$& 1.37 &$4.81\!\times\!10^{20}$ & 4.10&5.33&0.95\\
LuBi&$6.91\!\times\!10^{20}$& 1.91&$2.31\!\times\!10^{20}$& 1.85&$4.80\!\times\!10^{20}$&0.65&5.33&0.97\\
	\end{tabular}
	\end{ruledtabular}
	\caption{Parameters obtained from the analysis of magnetic field dependences of electrical conductivity and Hall conductivity with anisotropic multiband model.}
	\label{MR_ANALIZA_TABLE}
\end{table*}

These parameters allow us to estimate again the level of compensation of electrons and holes, expressed by the ratio $n_\alpha/(n_\beta+n_\delta)$ being equal to 0.95 for YBi and 0.97 for LuBi. Comparing them to analogous values from analysis of SdH oscillations (0.97 for YBi and 0.95 for LuBi), we conclude that electron-hole compensation is nearly perfect in both compounds. 
Small discrepancies between compensation values derived by different methods are most likely due to the approximations of Fermi pocket's shapes we made in our analyzes. 
\subsection*{Conclusions}
We investigated electron transport properties of high-quality single crystals of two compounds YBi and LuBi. The electronic structure that emerges from our results is almost identical for both compounds and points to their semimetallic character with nearly perfect compensation of electron and hole carriers. 
We found that low-temperature field-induced resistivity plateau could be interpreted in  terms of Kohler scaling with the main parameter confirming good compensation. 
This outcome is strengthened by our electronic structure calculations and analysis  of Shubnikov-de Haas oscillations revealing Fermi surfaces that consist of two hole pockets and a triplicate electron pocket. 
The multiband anisotropic model of electronic transport describes very well the experimental results of $\sigma_{xx}(B)$ and $\sigma_{xy}(B)$ for both compounds.
Therefore, our experimental results confirmed that prominent magnetotransport properties of YBi and LuBi could be explained without invoking nontrivial topology of electronic bands. 

Electronic structure calculations showed that band inversion exists in both compounds, but plausible Dirac points could appear about 0.5\,eV below the Fermi level (that is about twice as deep as in LaSb or LaBi \hspace{5pt}\cite{Tafti2016a}). There is also considerable $d-p\,$-orbital mixing of electron states visible in the same region. How such structures would influence magnetotransport of a semimetal remains an open question. 

The mobilities, of both electrons and holes, are considerably larger in YBi than in LuBi (Table~\ref{MR_ANALIZA_TABLE}), which is reflected in almost four times smaller residual resistivity of the former compound, and consequently leads to its three times larger magnetoresistance. But the band structure region where important orbital mixing occurs differs very little between YBi and LuBi (cf Fig.~\ref{el_struct}). This suggests that $d-p\,$-orbital mixing is not the predominant mechanism in magnetoresistance of these two compounds.

A scenario of mobility mismatch between electron and hole bands, proposed recently to explain reduced $M\!R$ in LaAs \hspace{5pt}\cite{Yang2017c}, does not seem appropriate for LuBi because its mobilities of holes and electrons differ very little, and the Hall coefficient is over two orders of magnitude smaller than in LaAs (for which the large Hall coefficient reflected strong mismatch of mobilities)\hspace{5pt}\cite{Yang2017c}. In the Supplemental Material we show also how YBi and LuBi follow $M\!R\propto RRR^2$ behavior, in common with several other monopnictide samples, but in contrast to LaAs \hspace{5pt}\cite{SupplMat}.     

Future research with the ARPES technique would be very helpful in making the final conclusion on the hypothetical presence of topologically nontrivial electronic states in YBi and LuBi.       
\subsection*{Acknowledgements}
This research was financially supported by the National Science Centre of Poland, grant no. 2015/18/A/ST3/00057. 
The band structure was calculated at the Wrocław Centre for Networking and Supercomputing, grant no. 359.
P. Swatek work at Ames Laboratory was supported by the US Department of Energy, Office of Science, Basic Energy Sciences, Materials Science and Engineering Division. Ames Laboratory is operated for the US Department of Energy by the Iowa State University under contract no. DE-AC02-07CH11358.
%
%
\large{\bf SUPPLEMENTAL MATERIAL}\vspace{-0.8cm}
\subsection*{Multiband models of magnetotransport}
\vspace{-0.3cm}
Analyzing magnetotransport of YSb, Xu et al. stressed inadequacy of isotropic multiband model to the properties of a system with anisotropic Fermi pockets \hspace{5pt}\cite{s_Xu2017f}. On the other hand in three papers devoted to LaBi \hspace{5pt}\cite{s_Tafti2016a,s_Kumar2016,s_Sun2016a}, their authors have used effective two-band model, neglecting the anisotropy of electron Fermi pocket centered at X-point, and found it satisfactory. We decided to compare how these both models work for YBi and LuBi. 
We used the same analysis as proposed by  Xu et al. \hspace{5pt}\cite{s_Xu2017f}:\\ 
magnetoconductivity due to a Fermi pocket, $p$, when magnetic field $B$ is applied along $z$-axis and electrical current is flowing along $x$-axis, is described by a tensor:
\begin{equation}
\hat{\sigma}^p =
\begin{pmatrix*}[r]
\sigma^p_{xx} & \sigma^p_{xy}\\
-\sigma^p_{xy} & \sigma^p_{yy}
\end{pmatrix*},
\label{sigma_tensor}
\end{equation}
with the components:
\begin{equation}
\begin{matrix*}[l]
\sigma^p_{xx} = e\,n_p\mu^p_x/[1+\mu^p_x\mu^p_yB^2]\\
\sigma^p_{yy} = e\,n_p\mu^p_y/[1+\mu^p_x\mu^p_yB^2]\\
\sigma^p_{xy} = e\,n_p\mu^p_x\mu^p_yB/[1+\mu^p_x\mu^p_yB^2],
\end{matrix*}
\label{sigma_tensor_components}
\end{equation}
where $n_p$ stand for carrier concentrations, $\mu^p_x$ and $\mu^p_y$ are two first diagonal components of mobility tensor for a Fermi pocket $p$, and $e$ is elementary charge. 

The hole pockets, centered at $\Gamma$ point have the cubic symmetry ($m\bar3m$ point group), and mobilities of the holes are isotropic, i.e. $\mu^p_x=\mu^p_y=\mu^p_z(\equiv\mu^p)$ for $p={\beta,\delta}$. 
Therefore: 
\begin{equation}
\begin{matrix*}[l]
\sigma_{xx}^p=e\,n_p\mu^p/[1+(\mu^pB)^2]\\
\sigma_{xy}^p=e\,n_p(\mu^p)^2B/[1+(\mu^pB)^2].
\end{matrix*} \quad{\rm for}\,p={\!\beta,\delta}
\label{sigma_tensor_components_holes}\end{equation}
Triplicate electron pocket consists of symmetry-equivalent parts of almost ellipsoidal shape, with long axes along main crystallographic axes. Each of them is centered at an X-point and therefore has point symmetry $4/mmm$. We distinguish them with symbols: $\alpha_x, \alpha_y$ and $\alpha_z$.  
Therefore mobility tensors for electrons of these sub-pockets can be written as: 
\begin{eqnarray}\nonumber
\hat{\mu}^{\alpha_x}=
\begin{pmatrix}
\mu_\parallel & 0 & 0 \\
0 & \mu_\bot & 0 \\
0 & 0 & \mu_\bot  
\end{pmatrix},\qquad\\
\hat{\mu}^{\alpha_y}=
\begin{pmatrix}
\mu_\bot & 0 & 0 \\
0 & \mu_\parallel & 0 \\
0 & 0 & \mu_\bot  
\end{pmatrix},\qquad\\ 
\hat{\mu}^{\alpha_z}=
\begin{pmatrix}
\mu_\bot & 0 & 0 \\
0 & \mu_\bot & 0 \\
0 & 0 & \mu_\parallel  
\end{pmatrix}.\qquad\nonumber 
\label{mu_tensors_electrons}
\end{eqnarray} 
We can express the ratio of independent components of these mobilities as: $\mu_{\bot}/\mu_{\parallel}=\kappa$. This parameter is equivalent of ($k_{F}^{\parallel}/k_{F}^{\bot})^2$, where $k_{F}^{\parallel}$ and $k_{F}^{\bot}$ are the Fermi wave vectors for each $\alpha_i$ pocket, parallel and perpendicular to its 4-fold symmetry axis, respectively. 

Thus, the symmetry reduces the number of parameters, and using Eq.~\ref{sigma_tensor_components}, the components of $\hat{\sigma}$ for the Fermi pockets $\alpha_x, \alpha_y$ and $\alpha_z$ can be written as:
\begin{equation}
\begin{matrix*}[l]
\sigma_{xx}^{\alpha_x}=\sigma_{yy}^{\alpha_y}=
e\,(n_{\alpha}/3)\mu_{\bot}/[\kappa+(\mu_{\bot}B)^2],\\\\
\sigma_{xx}^{\alpha_y}=\sigma_{yy}^{\alpha_x}=\sigma_{xx}^{\alpha_z}=\sigma_{yy}^{\alpha_z}= e\,(n_{\alpha}/3)\mu_{\bot}\kappa/[\kappa+(\mu_{\bot}B)^2],\\\\
\sigma_{xy}^{\alpha_x}=\sigma_{xy}^{\alpha_y}= 
e\,(n_{\alpha}/3)\mu_{\bot}^{2}B/[\kappa+(\mu_{\bot}B)^2],\\\\
\sigma_{xy}^{\alpha_z}=e\,(n_{\alpha}/3)\mu_{\bot}^{2}B/[1+(\mu_{\bot}B)^2].
\end{matrix*}
\label{sigma_tensor_components_electrons}
\end{equation}
Now total conductivity components are:
\begin{equation}
\sigma_{ij}=\sum_{p=\alpha_x, \alpha_y, \alpha_z, \beta, \delta}\sigma_{ij}^p. \label{total_sigma_components}
\end{equation}
From Eqs.~\ref{sigma_tensor_components_holes} and~\ref{sigma_tensor_components_electrons} it is obvious that $\sigma_{xx}=\sigma_{yy}$, which reflects the cubic crystal symmetry.
\begin{figure}[t]
	\includegraphics[width=8cm]{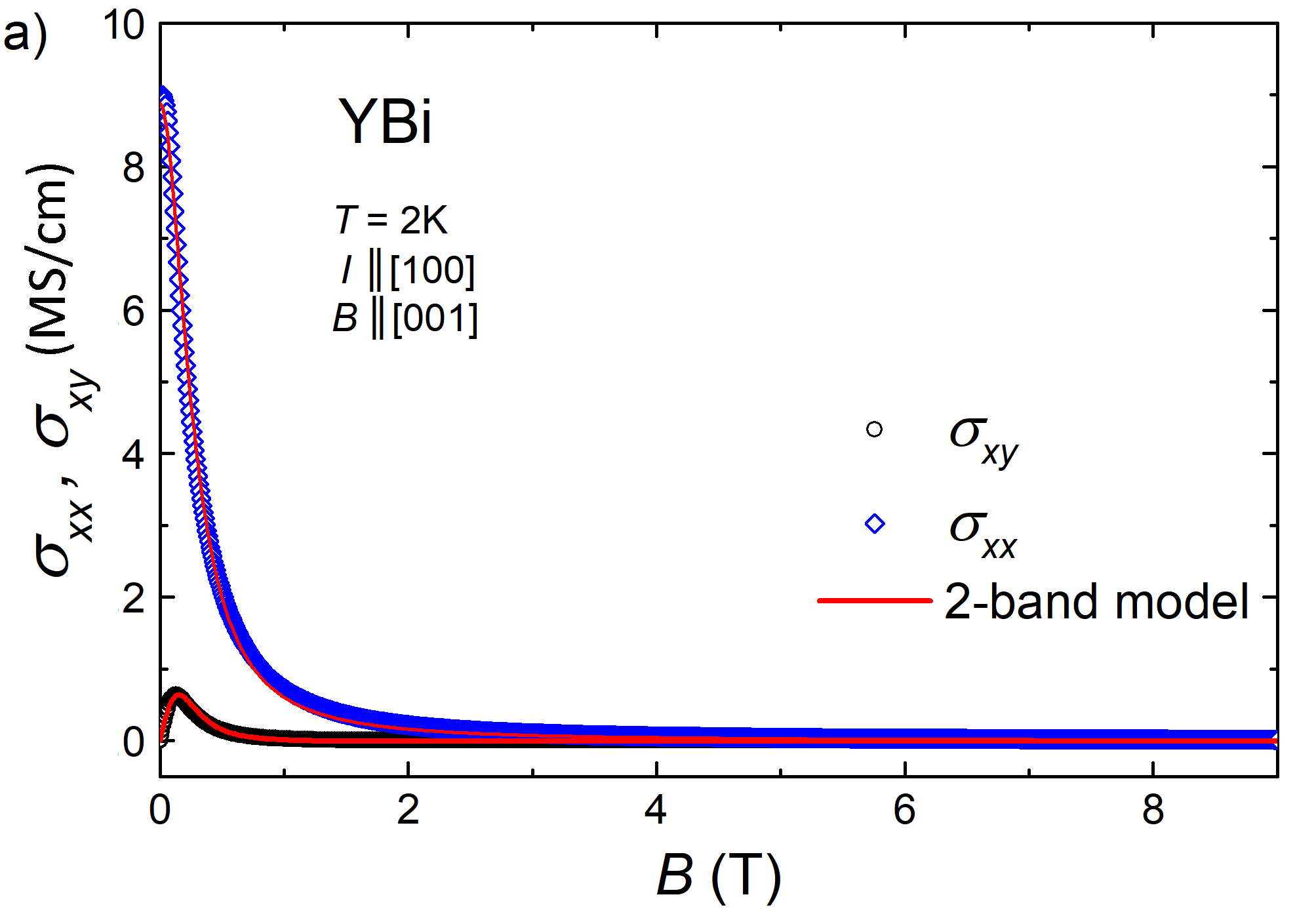}
   \includegraphics[width=8cm]{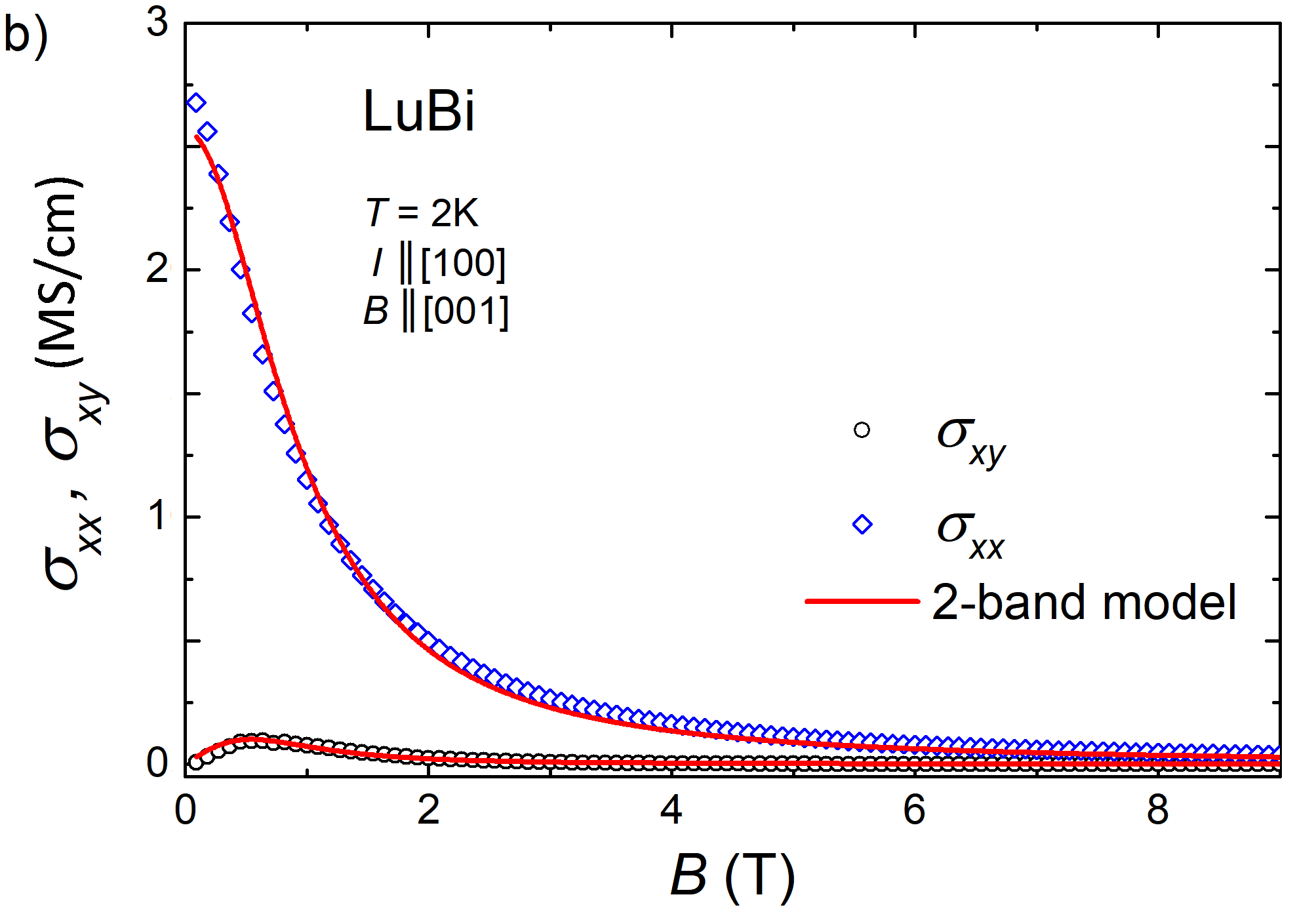}
\raggedright\textbf{FIG. S1} Magnetic field dependence of components of conductivity tensor $\sigma_{xx}$ and $\sigma_{xy}$, for YBi (a) and LuBi (b). Solid red lines represent  fitted with the effective two-band model.  
		\label{2b_sigma_fits}
\end{figure}

We fitted simultaneously both $\sigma_{xx}$ and $\sigma_{xy}$ of Eq.~\ref{total_sigma_components} to $\sigma_{xx}(B)$ and $\sigma_{xy}(B)$ data recorded at $T=2$\,K, with shared parameters. 
Both fitted functions have 7 parameters: $n_\alpha, \mu_\bot, n_\beta, \mu_\beta, n_\delta, \mu_\delta$ and $\kappa$, which are collected in Table~S1, together with $n_e/n_h$ ratios, and adjusted $R^2$ parameter reflecting the quality of the fit \hspace{5pt}\cite{OriginHelp}. 

We tested also effective two-band model, represented by Eq.~\ref{total_sigma_components} but after fixing values of $\kappa$ to 1 (assuming isotropic electron band) and $n_\delta$ to zero (leaving only one effective hole band). 
Such fits are shown in Fig.~S1 for data collected for both YBi and LuBi and can be compared to those made with anisotropic-three-band model shown in Fig.~9 of the paper. 
The effective two-band model yields worse fits than the anisotropic three-band model as seen in Figures and indicated by $R^2$ values given in Table~S1. 
Also the values of carrier concentrations and $n_e/n_h$ ratios from the anisotropic-three-band model much better correspond to those obtained from our analysis of SdH oscillations (cf. Table I of the paper), than these yielded by the two-band model.  
\begin{table*}[h]
\begin{ruledtabular}
	\begin{tabular}{ll *{9}{c}} 
\multicolumn{2}{l}{Compound}&
$n_\alpha$&$\mu_\bot$&$n_\beta$&$\mu_\beta$&$n_\delta$&$\mu_\delta$& $\kappa$&$n_e/n_h$&$R^2$\\
&model&$(\rm{cm}^{-3})$&(m$^2$/(V\!s))&$(\rm{cm}^{-3})$&(m$^2$/(V\!s))&$(\rm{cm}^{-3})$&(m$^2$/(V\!s))&$~~$ \\\colrule
YBi\\
& 3-band&~~$6.88\!\times\!10^{20}$& 6.92&$2.37\!\times\!10^{20}$& 1.37 &$4.81\!\times\!10^{20}$ & 4.10&5.33&0.95&0.9996\\
  &2-band&~~$7.36\!\times\!10^{20}$&4.16&$7.55\!\times\!10^{20}$&3.28&--&--&1 &0.97&0.9977\\ \hline 
LuBi\\
&3-band&~~$6.91\!\times\!10^{20}$& 1.91&$2.31\!\times\!10^{20}$& 1.85&$4.80\!\times\!10^{20}$&0.65&5.33&0.97&0.9999\\
   &2-band&~~$7.50\!\times\!10^{20}$ &1.13&$7.51\!\times\!10^{20}$ &1.00&--&--&1&1.00&0.9970\\
	\end{tabular}
\end{ruledtabular}\\\vspace{4mm}
\raggedright{\textbf{TABLE S1} Comparison of parameters obtained from the simultaneous fitting of magnetic field dependences of electrical conductivity tensor components $\sigma_{xx}$ and $\sigma_{xy}$ with the anisotropic three-band model and with the effective two-band model ($\kappa$~is fixed at 1 in 2-band model).}
	\label{MR_COMPARE}
\end{table*} 
\vspace{-1.2cm}
\subsection*{MR dependence on RRR$^2$}
\vspace{-0.3cm}
In Figure S2 we plotted magnetoresistance (MR) values versus square of residual-resistivity ratios (RRR$^2$) collected for several monopnictide samples described in different papers: YSb \hspace{5pt}\cite{s_Pavlosiuk2016f,s_Xu2017f}, LuSb \hspace{5pt}\cite{s_Pavlosiuk2017a}, LaSb \hspace{5pt}\cite{s_Tafti2016a}, LaBi \hspace{5pt}\cite{s_Tafti2016a,s_Yang2017c}, LaAs \hspace{5pt}\cite{s_Yang2017c}, as well as for our samples of YBi and LuBi. 
\begin{figure}[h]
\includegraphics[width=8cm]{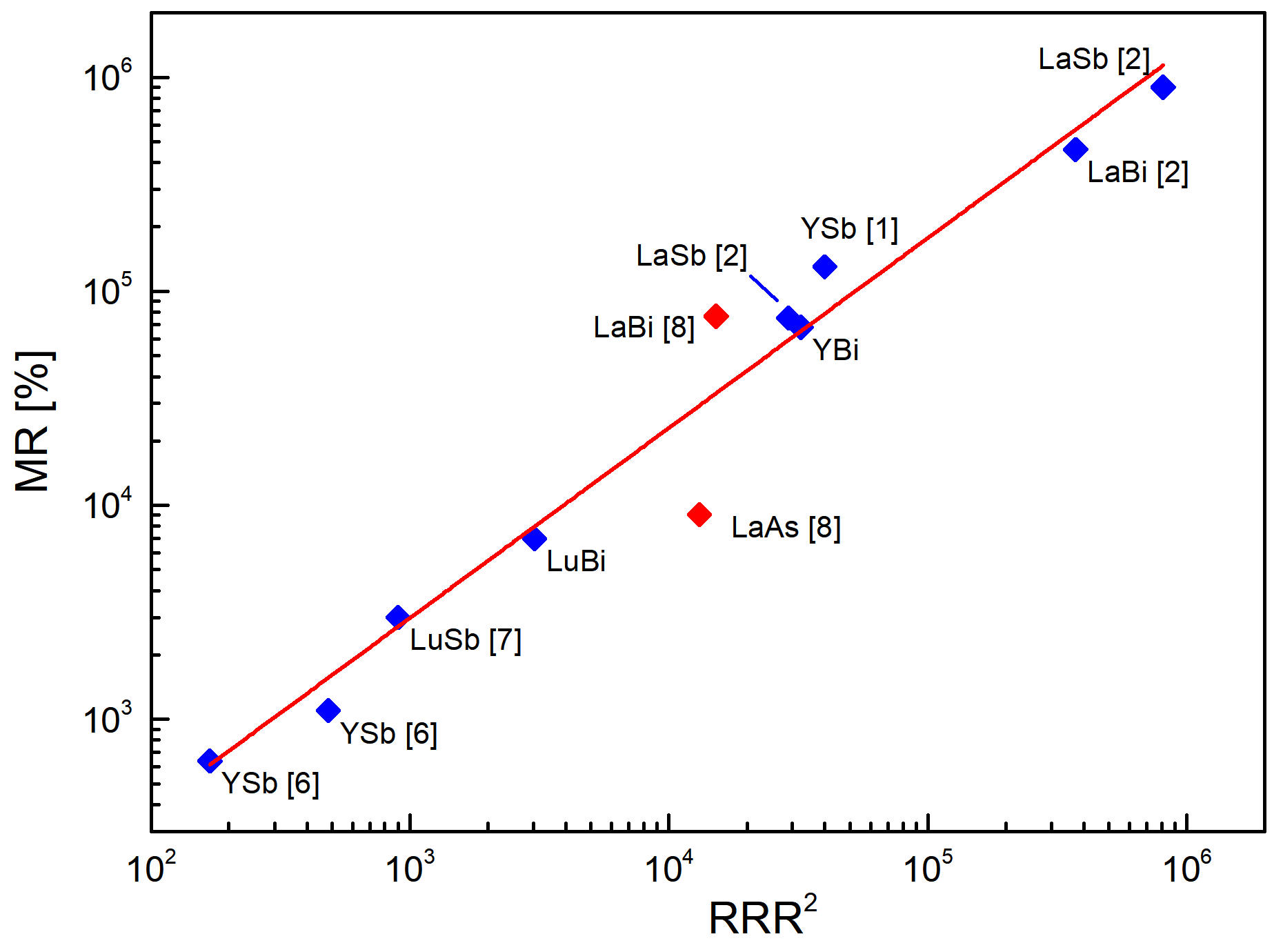}\\
\raggedright\textbf{FIG. S2} MR (at 2\,K and 9\,T) plotted versus RRR$^2$ for different monopnictide samples.
\end{figure}
Common MR$\propto$RRR$^2$ dependence is very closely followed for all the samples except LaBi and LaAs of Ref.~\onlinecite{s_Yang2017c}. For LaAs it has been argued that MR can be significantly reduced by strong mismatch of electron and hole mobilities \hspace{5pt}\cite{s_Yang2017c}. Since for both YBi and LuBi MR$\propto$RRR$^2$ and their mobilities of electrons and holes are not very different from each other, a scenario of mobility mismatch can be dismissed. 
%

\end{document}